\documentclass[aps,11pt,pra,showpacs,superscriptaddress,preprint]{revtex4}

\usepackage[T1]{fontenc} 
\usepackage{slashed}
\usepackage[utf8x]{inputenc}
\usepackage{color}
\usepackage[normalem]{ulem}
\usepackage{bm}
\usepackage{graphicx}

\begin{document}

\title{Searching for Doubly Charged Leptons at Present and Future Colliders}

\author{\textsc{S. Biondini}}
\affiliation{Physik Department, Technische Universit\"at M\"uenchen - Garching, Germany }

\author{\textsc{O. Panella}}
\affiliation{Istituto Nazionale di Fisica Nucleare, Sezione di Perugia, Perugia, Italy}

\author{\textsc{G. Pancheri}}
\affiliation{Laboratori Nazionali di Frascati - Roma, Italy	}

\author{\textsc{Y. N. Srivastava}}
\affiliation{Dipartimento di Fisica, Universit\`{a} degli Studi di Perugia, Perugia, Italy}

\author{\textsc{L. Fan\`o}}
\affiliation{Dipartimento di Fisica, Universit\`{a} degli Studi di Perugia, Perugia, Italy}

%\date{December 1, 2008}
\date{\today}

\begin{abstract}
The production at the LHC of exotic excited leptons of charge $Q = +2e$ is considered. Such
states are predicted in composite models with extended isospin multiplets ($I_{W}=1$ and
$I_{W}=3/2$). The coupling among these doubly charged leptons and Standard Model fermions may occurs either via gauge or contact interactions. In the former case the decay channels are more constrained. We study the production cross section at the LHC of $L^{++}$ ($pp \rightarrow L^{++} \, \ell^{-}$) and focus on the leptonic signature deriving from the subsequent decays  $L^{++} \rightarrow W^{+} \ell^{+} \rightarrow \ell^{+} \, \ell^{+} \, \nu_l $. 
The invariant mass distribution of the like-sign dilepton exhibits a sharp end-point corresponding to
excited doubly charged lepton mass $m^{*}$. A preliminary study for the production of doubly charged leptons at the future linear colliders, by considering the process $e^{-} e^{-} \rightarrow L_{e}^{--} \, \nu_{e}$, is carried out. Both the contact and gauge interaction mechanisms are investigated and compared. 
\end{abstract}

\maketitle

\section{Introduction}
It is typical of particle physics to reduce the number of elementary building blocks to understand how Nature works. Nevertheless the number of fundamental particles grew up to the present scheme with three generations of quarks and leptons. Moreover the heavier fermions decay into the lighter ones showing a rather typical dynamics of composite objects.  
We have three patterns of quarks and leptons organized in growing masses but sharing all remaining features
(charge, weak isospin, color). One may ask for a deeper and more satisfactory description of such evidences. A possible explanation for the replication of fermionic generations could be that they are not truly
fundamental particles but instead composite states of some yet unknown constituents. The possibility of a
further level of compositeness has been investigated phenomenologically for quite some time
\cite{ELP}, \cite{CaMaSr}.
We aim to emphasize a particular aspect of compositeness: the weak isospin invariance \cite{YN}. In this framework, the usual singlet ($I_{W}=0$) and doublet ($I_{W}=1/2$) isospin values are extended to include $I_{W}=0$ and $I_{W}=3/2$. Hence, multiplets (triplets and quartets) appear that contain exotic doubly charged leptons of charge $Q = +2e$ and exotic quark states of charge $Q = +(5/3)e$. Doubly charged leptons are also predicted within other models where they are stable particles \cite{stephan}. Bound states involving doubly charged leptons are considered as viable candidates for dark matter \cite{khol}. These exotic particles may generate interesting signatures to be searched for at the Large Hadron Collider (LHC). The energies involved at LHC are the highest ever reached so far. Thus, these new states might be produced and observed soon. The linear collider environment could be interesting as well. In particular, the possibility of same sign lepton beams, for example $e^{-} e^{-}$, provides a rather suitable setting for the experimental search of doubly charged particles. Despite a smaller nominal energy in the center of mass, at variance with hadron colliders, in a linear lepton collider the QCD background is strongly reduced. 

\section{Description of the models}
In early hadron physics, much progress was made by using symmetry arguments, namely strong isospin invariance. Indeed it was possible to discuss some patterns of baryon and meson resonances even when quarks and gluons were still unknown. On the same footing, as a great number of strong resonant low energy $\mathcal{O}(1)$ GeV states which were discovered, we may expect something similar in the electroweak sector, of course at much higher energies. The Higgs vacuum expectation value parameter, $v \simeq 238$ GeV, ought to play the role of the energy scale for possible fermion resonances, such that an expectation of some new physics at $\mathcal{O}(1)$ TeV scale seems natural. According to this, weak isospin ($I_{W}$) spectroscopy could reveal the quantum numbers of excited fermions without reference to any explicit internal dynamics of the building blocks.
To start with, all SM fermions belong to isospin doublets or singlets, namely $I_{W}=0$ and $I_{W}=1/2$, and the electroweak bosons have instead $I_{W}=0$ and $I_{W}=1$. Thus, only fermionic excited states with $I_{W} \leq 3/2$ can be predicted if one only uses the light SM fermions and electroweak gauge bosons. In order to compute the production and decays of these excited fermions, we need to define their couplings to Standard Model fermions and gauge bosons. The rules may be derived referring to weak isospin and hypercharge, Y. Since all the gauge fields have $Y=0$, excited fermions can only couple to Standard Model fermions with the same value of the hypercharge. Moreover, in order to preserve gauge invariance, we choose a transition current containing a $\sigma_{\mu \nu}$ term and not a single $\gamma_{\mu}$ (one may insert also the vector current but we do not consider the case). This last consideration automatically provides electromagnetic current conservation. 
The Lagrangian density  for $I_{W}=1$ is:
\begin{equation}
\label{lag1}
\mathcal{L}_{\rm{GI}}=\frac{gf_{1}}{m^{*}}\left( \bar{L} \, \sigma_{\mu \nu} \, \partial^{\nu} \, W^{\mu} \, \frac{1+\gamma^{5}}{2} \, \ell \right)  + h. c.
\end{equation}
while  for $I_{W}=3/2$ reads:
\begin{equation}
\label{lag32}
\mathcal{L}_{\rm{GI}}=C_{\left( \frac{3}{2}, M | 1,m; \frac{1}{2}, m'\right)} \frac{gf_{3}}{m^{*}}\left( \bar{L}_{M}\sigma_{\mu \nu} \, \partial^{\nu} \, W^{\mu} \, \frac{1-\gamma^{5}}{2} \, \ell_{m'}  \right)  + h.c.
\end{equation}
In the above equations, $L$ stands for the excited lepton spinor, $m^{*}$ is the excited fermion mass, and $f_{1}, f_{3}$ are dimension-less coupling constants, 
expected to be of order one and whose precise value can only be fixed through a specific choice of the compositeness model. The Clebsch-Gordon coefficients in \ref{lag32} account for the different combinations of the elements in the isospin multiplets, either SM or excited leptons.   

There is another complementary mechanism to produce the composite fermions in the effective field theory approach. One may consider standard four-fermion contact interactions by ignoring mediating gauge bosons (either Standard Model or new additional ones). In this case the Lagrangian reads:
\begin{equation}
\mathcal{L}_{\rm{CI}}=\left( \frac{g^{2}_{*}}{2\Lambda^{2}}\right) j^{\mu} j_{\mu} \, , 
\label{cont}
\end{equation}
where the current for the specific channel of interest, $e^{-} e^{-} \rightarrow L^{--}_{e} \, \nu_{e}$ is the following
\begin{equation}
j_{\mu}= \left( \frac{g^{2}_{*}}{2\Lambda^{2}}\right) \left[ \bar{\nu}_{e}(x) \gamma_{\mu} P_{L} e(x) + \bar{L}_{e}(x)\gamma_{\mu} P_{L} e(x) + h.c. \right]  \, .
\end{equation}   
All the fields correspond to the particles entering in the process: electrons in the initial state, doubly charged excited lepton and electron neutrino in the final state. 

In order to perform the numerical calculations of the production cross sections and the kinematic distributions, we need to implement our model in the CalcHEP event generator. We implemented the above Lagrangians in eqs. (\ref{lag1}-\ref{cont}) through FeynRules~\cite{Feyn}, a Mathematica~\cite{math} package that generates the Feynman rules of any given quantum filed theory model \cite{simo}, \cite{Leo}.

\section{Doubly charged leptons phenomenology}
\begin{figure}
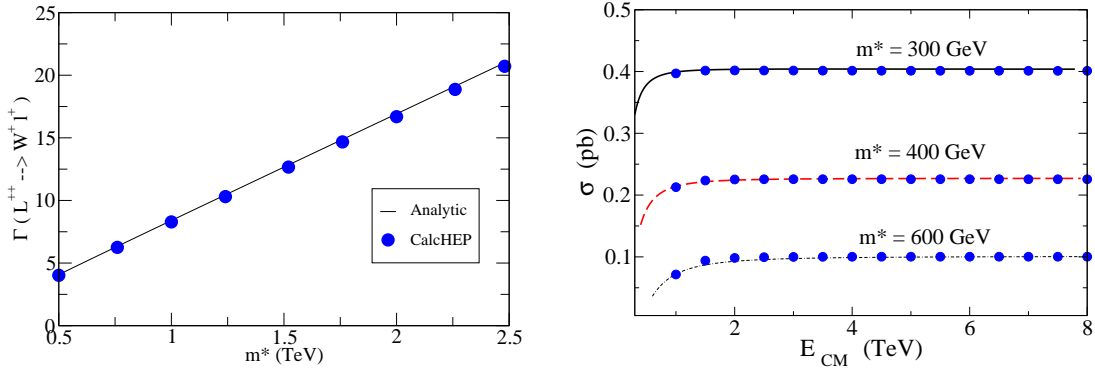

\centering
\vspace{0.2 cm}
\includegraphics[scale=0.87]{decay}
\hspace{0.4cm}
\includegraphics[scale=0.97]{IMP}
\caption{\label{fig1} In the left panel, the total decay width of the exotic lepton $L^{++}$ is shown as a  function of its mass $m^*$. We compare the analytical result with the CalcHEP output (blue dots).  The total parton cross section for $L^{++}$ (and $L^{--}$, there is no difference at the parton level) against the energy in center of mass frame is shown in the right panel. We plot the parton cross section for three different values of the doubly charged lepton mass. The analytical results are compared with CalcHEP output (blue dots).}
\end{figure}

In this section we consider the phenomenology of the doubly charged lepton in the case of gauge interactions and within the LHC framework \cite{simo}. 
Provided that the doubly charged $L^{--}$ and $L^{++}$ interact with the Standard Model fermions only via the gauge interactions, one can easily compute the decay width of these exotic states.
Indeed the only available decay channel of the doubly charged lepton is $L^{++} \to W^+ \ell^+$. Then the branching ratio is exactly one: $\mathcal{B} (L^{++} \to W^+ \ell^+)=1$.  The analytic expression  of the total decay width reads as follows:
\begin{equation}
\label{decaywidth}
\Gamma_{L^{++}}= \Gamma (L^{++} \rightarrow W^{+} \, \ell^{+}) = \left( \frac{f}{\sin \theta_{W}} \right)^{2} \alpha_{\rm{QED}} \frac{m^{*}}{8} \left( 2+ \frac{M^{2}_{W}}{m^{*2}} \right) \left( 1- \frac{M^{2}_{W}}{m^{*2}} \right)^{2}   
\end{equation}
According to the large mass for the excited states with respct to Standard Model particles, we can use the approximation $M_{W} \ll m^{*}$. In this limit, the decay width increases linearly with the mass i.e. $\Gamma= \kappa m^{*} $ as shown in Fig.\ref{fig1}. 

\begin{figure}
\center
\includegraphics[scale=0.69]{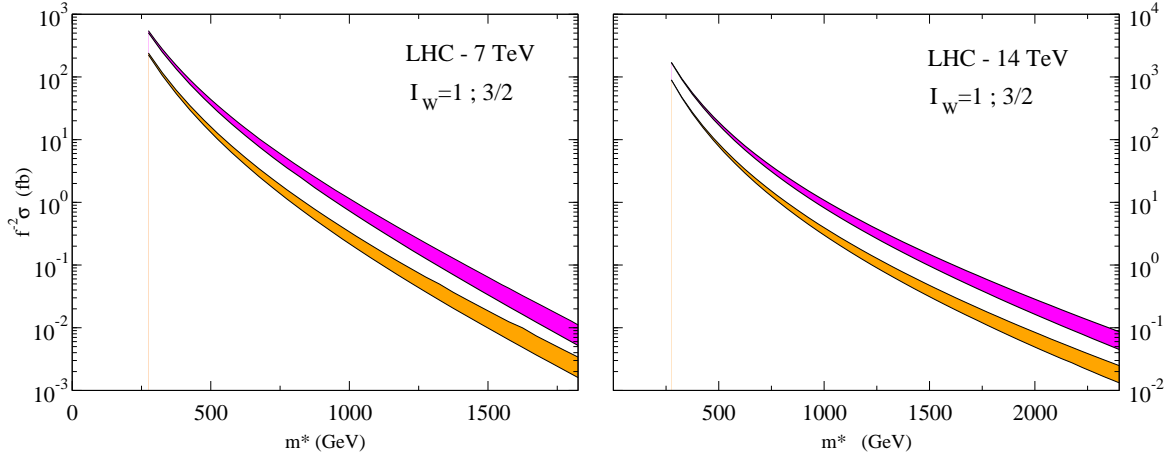}
\caption{\label{fig2} The total production cross section is shown for the LHC run at 7 TeV (left panel) and 14 TeV (right panel). The production cross section for the positively doubly charged leptons (in magenta) is bigger than the one for the negatively doubly charged leptons (in orange). The uncertainty bands (magenta and orange) correspond to the running of the  factorization and renormalization scale from $Q=M_W$ up to $Q=m_*$.
The total production cross section is independent of different isospin values, $I_{W}=1$ and $I_{W}=3/2$, due to the structure of the parton cross section.}
\end{figure}
Following the convention and notation of \cite{YN}, we write down the parton cross section for both $I_{W}=1$ and $I_{W}=3/2$. We write the expressions for $L^{++}$ by using the Maldestam variables:
   
\begin{eqnarray}
\left( \frac{d\hat{\sigma}}{d\hat{t}}\right)_{u\bar{d} \rightarrow L^{++} \ell^-}&=&
\frac{1}{4m^{*2}s^{2}} \frac{f^{2}}{12 \pi} \frac{s}{(s-M^{2}_{W})^{2}+(M_{W} \Gamma_{W})^{2}}\\&&\phantom{xxxx} \left\lbrace  \left( \frac{g^{2}}{4} \right) \left[ m^{*2}(s-m^{*2}) + 2ut \right] \pm 2 \left( -\frac{g^{2}}{8} \right) m^{*2} (t-u) \right\rbrace  \nonumber
\label{p1}
\end{eqnarray}
where $f$ is the unknown coupling related to the different isospin multiplets.
There is a difference between the two cases we want to stress. The positive sign in eq.~\ref{p1} must be used for $I_{W}=1$ while the negative sign must be used for $I_{W}=3/2$.
One may easily obtain the expression for the total parton cross section integrating all over the angular variables.

In order to get the production cross section, we exploit the CalcHEP numerical session. We have used the CTEQ6m (in CalcHEP library) parton distribution functions and we summarize the results in Fig.\ref{fig2}. We observe that the production cross section for positively doubly charged leptons is favoured with respect to the negatively charged ones. This is due to the parton proton content. 

\section{Invariant mass distribution}
\begin{figure}
\centering
\includegraphics[scale=0.92]{imd}
\hspace{0.3 cm}
\includegraphics[scale=0.38]{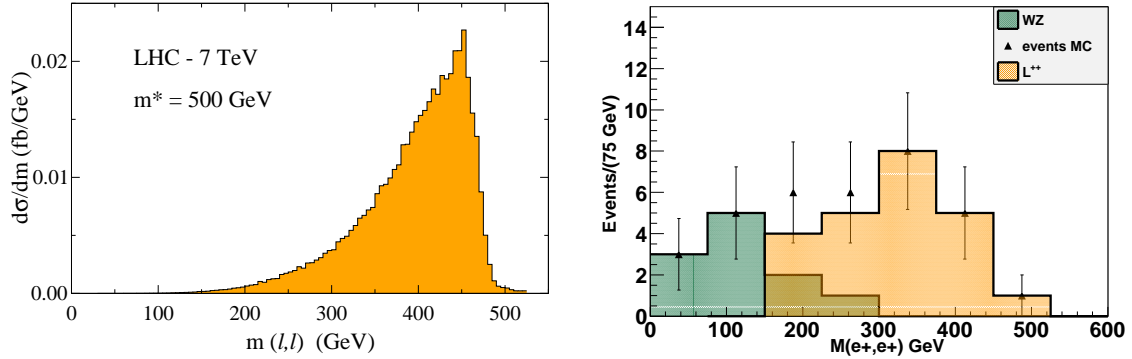}
\caption{ \label{fig3} The invariant mass distribution of the like-sign leptons in the final state is shown. In the left panel, we plot the invariant mass distribution at the level of the CalcHEP generator for the signal only. The end-point shape is well reproduced. In the right panel, we show both the signal (filled orange) and background (filled green) invariant mass distribution after the detection simulation (run at $\sqrt{s}=7$ TeV) for an integrated luminosity of $L=10$ fb$^{-1}$. A value of $m^{*}=500$ GeV is assumed for the mass of the doubly charged lepton.}
\end{figure}

We consider the doubly charged leptons produced in proton-proton collisions, namely the process $pp \rightarrow L^{++} \, \ell^{-}$. Then, we focus on the final state particle set determined by the following decays: $L^{++} \rightarrow W^{+} \, \ell^{+} \rightarrow \ell^{+} \, \ell^{+} \, \bar{\nu}_{\ell}$. At the end one has $pp \rightarrow \ell^{-} \, \ell^{+} \, \ell^{+} \, \bar{\nu}_{\ell}$. In this case, as suggested in~\cite{MR} one should observe a like-sign dilepton mass invariant distribution with a sharp end point at $m^{*}$, which is also rather close to the maximum of the distribution. We get the same for the signal under consideration as one may see in the left panel in Fig.~\ref{fig3}. We consider the standard model irreducible background. We compare the invariant mass distributions generated by the signal and the background at the level of the CalcHEP generator.

Finally we provide a more realistic description of the final state event at LHC via a fast simulation of the detector based on the PGS software \cite{PGS}. The distributions of the main kinematic variables are, initially, obtained as CalcHEP output. Hence, they do not refer to typical reconstructed objects one deals  with in an actual experimental analysis. They are kind of ideal data. The Pretty Good Simulator~\cite{PGS} package allows to perform a reconstruction of the particles in the final state event. Indeed one can introduce the efficiency of the detectors such as trackers resolution, calorimeter geometrical acceptance and energy resolution. Here we focus on the invariant mass distribution only. The result for the signal and background reconstructed invariant mass distribution is shown in Fig.\ref{fig3}. 

\section{Production cross sections at linear colliders: preliminary results}
\begin{figure}
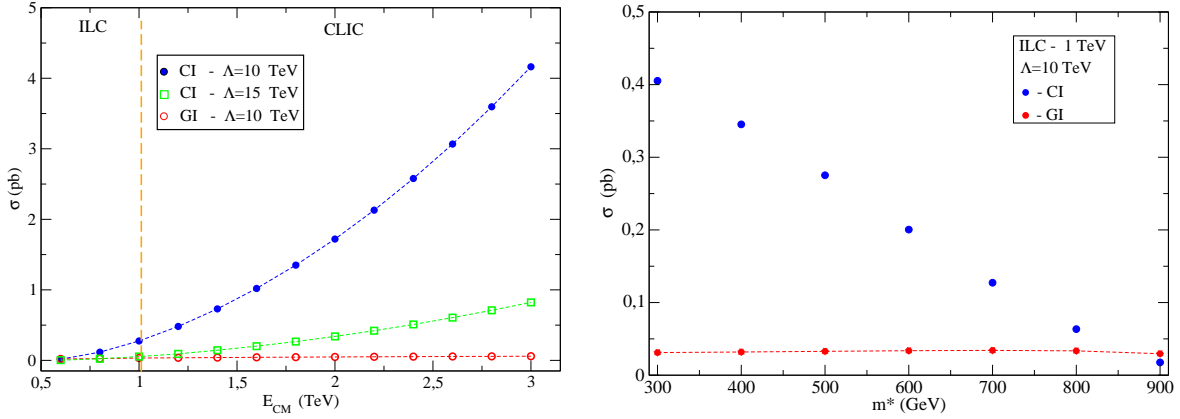

\centering
\includegraphics[scale=0.32]{cross500}
\hspace{0.25 cm}
\includegraphics[scale=0.32]{cross}
\caption{\label{fig4} We show the production cross section for the process $e^{-} e^{-} \rightarrow L_{e}^{--}  \nu_{e}$ within the framework of gauge and contact interactions. In the left panel, the production cross section for different values of the compositeness scale and production mechanisms are shown. The mass for the doubly charged lepton set to 500 GeV. In the right panel, we compare the cross sections versus the mass of the excited lepton at fixed energy in the center of mass reference frame (1 TeV).}
\end{figure}

In this section we report some preliminary results for the production of the doubly charged excited leptons at the future linear collider facilities. We consider both the International Linear Collider (ILC) and Compact Linear International Collider (CLIC). The latter is expected to reach higher energies up to 3 TeV with respect to the ILC (which is expected to achieve a maximum energy of 1 TeV). By exploiting lepton beams, a cleaner and low-multiplicity final state is realized. Indeed, the QCD background is strongly suppressed with respect to the hadron collisions where the beam remnants make the final state much more involved to study. Moreover, the option of $e^{-}e^{-}$ collisions has been put forward within the linear collider projects. Such possibility would be, of course, very interesting for the search of doubly charged leptons. Indeed, we can consider the production according to the channel $e^{-} e^{-} \rightarrow L_{e}^{--} \nu_{e}$, accounting for both the gauge and contact interaction mechanisms in eq.~(\ref{lag1}-\ref{cont}). Only the production of electronic doubly charged leptons may be considered since the present model does not include flavour-changing processes. We calculate the cross section by using the event generator CalcHEP. We obtain the production cross sections as a function of the mass of the excited doubly charged lepton $L_{e}$ and of the energy in the center mass reference frame. The results are shown in Fig.~\ref{fig4}.   
The contact interaction mechanism provide a higher cross section for the production cross section up to 900 GeV for the excited lepton mass. 

\section{Conclusions}
The phenomenology of doubly charged excited leptons at LHC has been discussed almost within the framework of gauge interactions. The parton and production cross section have been calculated, together with the decay width according to the only available channel $L^{++} \rightarrow \ell^{+} \, W^{+}$. Four-fermion contact interactions are briefly mentioned and they are the subject of a forthcoming study in \cite{Leo}. The invariant mass distribution of the like-sign lepton system has been studied in detail at the level of the CalcHEP generator. A fast simulation of a generic detector response to our signature has been implemented through the PGS software: experimental uncertainties and the smearing of the main kinematic variables are introduced.

We have briefly discussed some preliminary results for the production cross section of doubly charged leptons in the framework of the forthcoming linear colliders. Both gauge and four-fermion contact interaction have been included in the production mechanism.  
A more accurate analysis needs to be carried out in the framework of the linear collider facilities to make more quantitative predictions and comparisons with the LHC facility \cite{simo2}. 

\section{Acknowledgements}
While the LHC results presented in this work were also discussed in \cite{simo}, the LC findings are a preliminary account of a forthcoming study \cite{simo2}.
The speaker (S.B.) warmly thanks the organizers for the kind invitation to the LC13 conference.

\end{document}